\documentclass[preprint,12pt]{elsarticle}

\usepackage{amssymb}
\usepackage{makecell}
\usepackage{lineno}
\journal{}

\begin{document}

\begin{frontmatter}

%% Title, authors and addresses

%% use the tnoteref command within \title for footnotes;
%% use the tnotetext command for theassociated footnote;
%% use the fnref command within \author or \address for footnotes;
%% use the fntext command for theassociated footnote;
%% use the corref command within \author for corresponding author footnotes;
%% use the cortext command for theassociated footnote;
%% use the ead command for the email address,
%% and the form \ead[url] for the home page:
%% \title{Title\tnoteref{label1}}
%% \tnotetext[label1]{}
%% \author{Name\corref{cor1}\fnref{label2}}
%% \ead{email address}
%% \ead[url]{home page}
%% \fntext[label2]{}
%% \cortext[cor1]{}
%\affiliation{organization={},
%             addressline={},
%             city={},
%             postcode={},
%             state={},
%             country={}}
%% \fntext[label3]{}

\title{ Multi-physics integration platform MuPIF: Application for composite material design}

%% use optional labels to link authors explicitly to addresses:
%% \author[label1,label2]{}
%% \affiliation[label1]{organization={},
%%             addressline={},
%%             city={},
%%             postcode={},
%%             state={},
%%             country={}}
%%
%% \affiliation[label2]{organization={},
%%             addressline={},
%%             city={},
%%             postcode={},
%%             state={},
%%             country={}}

\author[inst1]{B. Patzák}
\author[inst1]{V. Šmilauer}
\author[inst1]{M. Horák}
\author[inst1]{S.  Šulc}
\author[inst1]{E. Dvořáková}

\affiliation[inst1]{organization={Czech Technical University in Prague,  Faculty of Civil Engineering, Department of Mechanics},
            addressline={Thákurova 7}, 
            city={Prague 6},
            postcode={166 29},
            country={Czech Republic}}

\begin{abstract}
%% Text of abstract
This paper presents the design of the MuPIF distributed, multi-physics simulation platform and its performance in the context of the H2020 Composelector project. The description of MuPIF’s model and data interfaces provides implementation and operational details that illustrate how MuPIF is a highly versatile and robust tool for various engineering applications. Its distributed design and the integration of models in a workflow allows MuPIF to simulate real industrial tasks. The design of a composite airplane frame for the Composelector project, focusing on innovative design of composite materials and structures, demonstrates MuPIF’s capabilities and ability to be integrated into Business Decision Support Systems.
\end{abstract}

%%Graphical abstract
%\begin{graphicalabstract}
%\includegraphics{grabs}
%\end{graphicalabstract}

%%Research highlights
%\begin{highlights}
%\item Research highlight 1
%\item Research highlight 2
%\end{highlights}

\begin{keyword}
multi-physics platform\sep integration platform\sep interoperability\sep object-oriented design\sep airplane frame
%% keywords here, in the form: keyword \sep keyword

%% PACS codes here, in the form: \PACS code \sep code

%% MSC codes here, in the form: \MSC code \sep code
%% or \MSC[2008] code \sep code (2000 is the default)

\end{keyword}

\end{frontmatter}

%% \linenumbers

%% main text
\section{Introduction}
\label{sec:Intro}
Composite materials such as fiber-reinforced materials, laminated composites, and functionally graded structures are omnipresent in everyday life. Their use is growing fast, with applications in many fields including aerospace, mechanical, and biomechanical engineering, to name but a few. Tailoring and optimizing the properties of composites such as stiffness, weight, and thermal conductivity for specific applications leads to increased performance and significant cost savings. The ability to create a digital “virtual twin” by means of modeling of a real problem makes the selection and optimization of new products faster and cheaper than traditional approaches based on experience and prototype evaluation. This opens up new opportunities for exploring the untapped potential of new materials.
Consequently, mathematical modeling is becoming essential to the design and optimization of a composite material. However, individual composite material components exhibit different features at varying time and length scales, a problem referred to as the “tyranny of scales” \cite{[1]}. Material properties are also highly influenced by different manufacturing processes.

Therefore, the development of reliable multiscale and multiphysics models for the simulation of composites materials, including the manufacturing process, is a very complex and challenging task. The development of such tools would constitute a significant step beyond the current state-of-the-art in many fields, but it would require enormous effort. Today, many advanced single-scale models are available. Therefore, from the practical point of view, the re-use of the existing tools and their integration is preferred over the development of new models. However, the individual existing models use different numerical methods and rely on specific data formats, which present another level of challenge and complexity, from the software engineering perspective, in particular \cite{[2]}.
 
Thus, there is a strong need for integration platforms enabling documented, automated, and repeatable inter-operable integration of existing models and data sources into automated simulation workflows. Interoperability, by definition, implies standards. Traditional approaches build on syntactic interoperability, implying specific communication protocols and data conversions. Today, the most attractive approaches are based on semantic interoperability, where data and models are brought together with their meanings, which then allows for automated machine interpretation, translation, data manipulation, and verification \cite{[3]}. Many of the existing multiphysics simulation toolboxes enable such integration of models and can be used with external third-party modules. Such capabilities are found, e .g., in ADINA \cite{[4]}, MSC Nastran \cite{[5]}, ANSYS Multiphysics \cite{[6]}, Elmer \cite{[7]},  OpenFOAM \cite{[8]}, and COMSOL \cite{[9]}. Several independent integration platforms have also been developed over the past few decades, particularly for specific purposes and for solving problems in specific domains while having generic designs, including Moose \cite{[10]}, SIERRA \cite{[11]}, AixVipMap \cite{[12]}, Trilinos \cite{[13]}, among others. A comprehensive overview of such platforms can be found in \cite{[14]}.

Recently, several EU funded projects (DEEPEN \cite{[15]}, MoDeNa \cite{[16]}, NanoSim \cite{[17]}, SimPhoNy \cite{[18]}), as a part of the EU Multiscale Materials Modelling Cluster (EMM Cluster), have focused on developing open, integrated, and multi-purpose numerical nano-enabled design environments. The EMM Cluster aims at enabling knowledge exchange and  fostering the adoption of novel approaches for multiscale modeling, and it provides a platform for harmonizing standardization. In parallel, the European Material Modeling Council (EMMC) \cite{[3]} was set up as a community-driven, bottom-up action in order to connect all existing material modeling activities in Europe, focusing on establishing a common language and standards. EMMC’s activities have resulted in the definition for the classification of materials models, and a standardized description of materials modeling applications based on a system referred to as MODA (Materials Modeling Data) \cite{[19]}. The EMMC has also created the European Materials Modelling Ontology (EMMO) \cite{[20]}, \cite{[21]}, a basis for semantic interoperability.

In this paper, we describe the design of the MuPIF distributed, object-oriented simulation platform \cite{[22],[23]} developed in the context of two EU funded projects \cite{[16], [30]}. Its use and operation is illustrated here in a real example, the design of a composite airframe. We discuss the platform’s unique design and development since 2013, when the first paper introducing MuPIF was published \cite{[22]}. MuPIF is an open-source project developed in Python 3 and licensed under the LGPL license.
%%%%%%%%%%%%%%%%%%%%%%%%%%%%%%%%%%%%%%%%%%%%%%%%%%%%
\section{Design of MuPIF platform}
\begin{figure}[h]
\begin{center}
\centerline{\includegraphics[width=1.0 \linewidth]{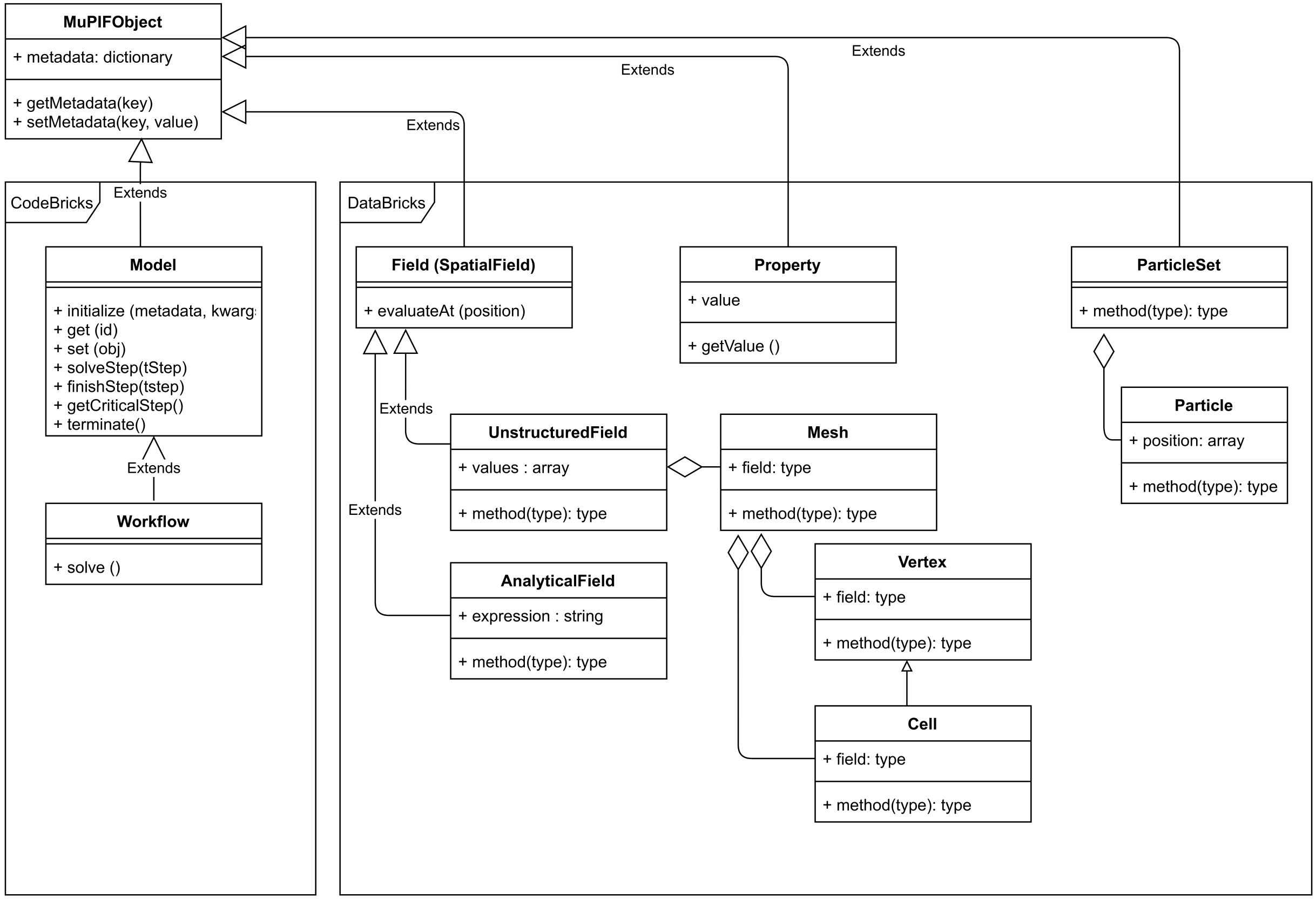}}
\caption{MuPIF object-oriented structure (simplified).} 
\label{Fig:MuPIF_OOStructure}
\end{center}
\end{figure}

MuPIF utilizes an object-oriented approach, with classes introduced to represent identified entities. These classes include, among others, individual simulation models and exchanged data types, see Figure \ref{Fig:MuPIF_OOStructure}.

The elegance of the object-oriented approach, with classes introduced to represent identified entities. These classes include, among others, individual simulation models and exchanged data types, see Figure \ref{Fig:MuPIF_OOStructure}. The elegance of the object-oriented approach lies in its ability to express similarities between classes using inheritance, a concept that allows one class to be derived from another using the inheritance of attributes and services of a parent class. A derived class can extend a parent class by i) adding new attributes or methods and/or ii) providing its implementation for inherited methods (specialization). In this way, a parent class can define an interface (in terms of defined services) common to all derived classes. The terms “abstract classes” refers to the top-level generic classes that define interfaces.
In MuPIF, generic abstract classes are introduced to represent common interfaces for sets of similar entities that facilitate communication as well as the manipulation of derived classes (and entities they represent) using the same interface. This concept is used to define a common interface for all models, represented by a “Model” class, described in detail in the next section. One of MuPIF’s crucial and distinct features is that abstract classes defining generic interfaces are also defined for data types. The instances of these classes represent data and are exchanged between models. The class representation of data types is established for relatively simple entities such as properties, but also for more complex entities such as spatial fields and microstructure representations. This feature allows all specific representations of particular data types on an abstract level to be handled using a generic interface declared by a corresponding abstract class. In this way, abstraction from a particular internal data representation of a data type, including storage and location, can take place. In turn, the models working with the data obtain required information from data objects using services, rather than obtaining them by interpreting raw data (which yields the data format dependence). One can think of abstract classes as representing data as “data bricks” with standardized connectors able to be used in their appropriate place in workflows to represent abstract data containers. In summary, the adopted design concept facilitates a true plug \& play architecture in which individual models as well as data representations can be plugged into simulation workflows and manipulated using  generic interfaces.  
MuPIF achieves interoperability with standardization of application and data component interfaces and it is not reliant on standardized data structures or protocols. Any existing data format can be plugged in and used transparently, provided the corresponding data interface is implemented.
All classes in MuPIF are derived from the top-level abstract class, \textit{MuPIFObject}, which declares a generic interface common to all classes. It specifically defines the services, allowing additional information about an object itself (metadata) to be attached. Metadata play an important role here, since they facilitate physical meaning and the origin of data, units, and so on. The provider must define some metadata while the platform automatically collects others. Each class defines metadata and metadata structure requirements by providing a metadata template in a standardized JSON format \cite{[24]}. Object metadata can be validated against the template to ensure consistency.
MuPIF is written in the Python 3 programming language, is platform independent, and supports different operating systems including Windows, Linux, Unix, and Mac. MuPIF can interface with closed source software, commercial code, and open-source code, as described in the following section.
%%%%%%%%%%%%%%%%%%%%%%%%%%%%%%%%%%%%%%%%%%%%%%%%%%%%%%%%%%%%%%%%%%%%%%%%%%
\section{MuPIF’s model interface}
The \textit{Model} interface facilitates the performance of data exchange with a model, typically a simulation tool implemented in the software, and steers its execution. Such data exchange methods allow for setting or getting a specific data object that can represent a simple data object such as property, or a more complex data object, like a spatial field. Individual applications are not supposed to interpret the raw data structures of data objects, but rather query data objects using standardized interfaces. The advantages of this concept are discussed in the next section.
The steering services make it possible to update the model state for a given time step, to report a critical time step, to check model status, and so on.

\begin{table}[]
\footnotesize
\begin{tabular}{|p{5.5cm}|p{7.5cm}|}
\hline
\textbf{Model’s method} & \textbf{Description}
\\ \hline
Init 			(inFile, 			workdir)                                & Constructor. 			Initializes the model.                                                                \\ \hline
getDataComponent(propID, 			time)                           & Returns 			data identified by its ID evaluated at given time.                                         \\ \hline
setDataComponent(self,component)                        & Registers 			the given data component in application.                                                 \\ \hline
solveStep(self, 			tStep, stageID=0) & Solves 			the problem for a given time step. tStep is object representing 			solution time step.      \\ \hline
isSolved(self)                                              & Returns 			true or false depending whether solution has completed when 			executed in the background. \\ \hline
wait(self)                                                  & Wait 			until solution is completed when executed in background.                                      \\ \hline
finishStep(self, 			tstep)                                  & Called 			after a global convergence within a solution time step.                                     \\ \hline
\end{tabular}
\caption{Model interface definition (simplified).}
\label{Tab:ModelInterfaceDefinition}
\end{table}

Table \ref{Tab:ModelInterfaceDefinition} lists the most important methods for the application interface. The design assumes individual models combined in sequential or loosely coupled workflows, see Figure 2.
Simulation workflows are composed of individual models. A simulation workflow itself can also be considered to be a (more complex) model. Workflows are also represented by classes derived from~the~\textit{Model}; thus, any workflow implements a \textit{Model} interface, which is another unique feature of the MuPIF platform. Thus, all models and workflows use the same generic interface and use workflows in other workflows like any different model, resulting in a hierarchy of simulation workflows established naturally.
In order to connect an existing simulation tool to the platform, one needs to implement the \textit{Model} interface for a specific tool. Implementation consists of creating a derived class from the \textit{Model} class while implementing the required methods. Actual implementation can be performed in a number of ways depending on the particular application.  In general, two different approaches are possible:
\begin{itemize}
    \item Direct approach: Based on direct communication with the application using its scripting or programming interface (requires linking the application as a library or using some inter-process communication mechanism).
    \item Indirect communication with a simulation tool, usually using input/output files: In this case, the application interface records all settings of parameters and variables. When a model update is requested, an input file is produced, a simulation tool is executed, and an output file is parsed for output parameters. Several such interfaces have been created to date, including both commercial and open-source simulation tools. Examples are provided later in this paper.
\end{itemize}

\begin{figure}[h]
\centering
\begin{tabular}{cc}
\includegraphics[width=0.2 \linewidth]{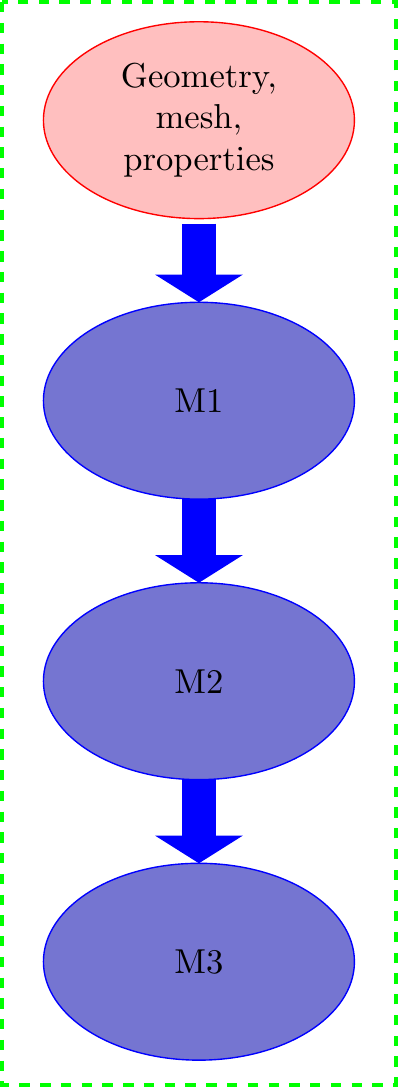}
&
\includegraphics[width=0.465 \linewidth]{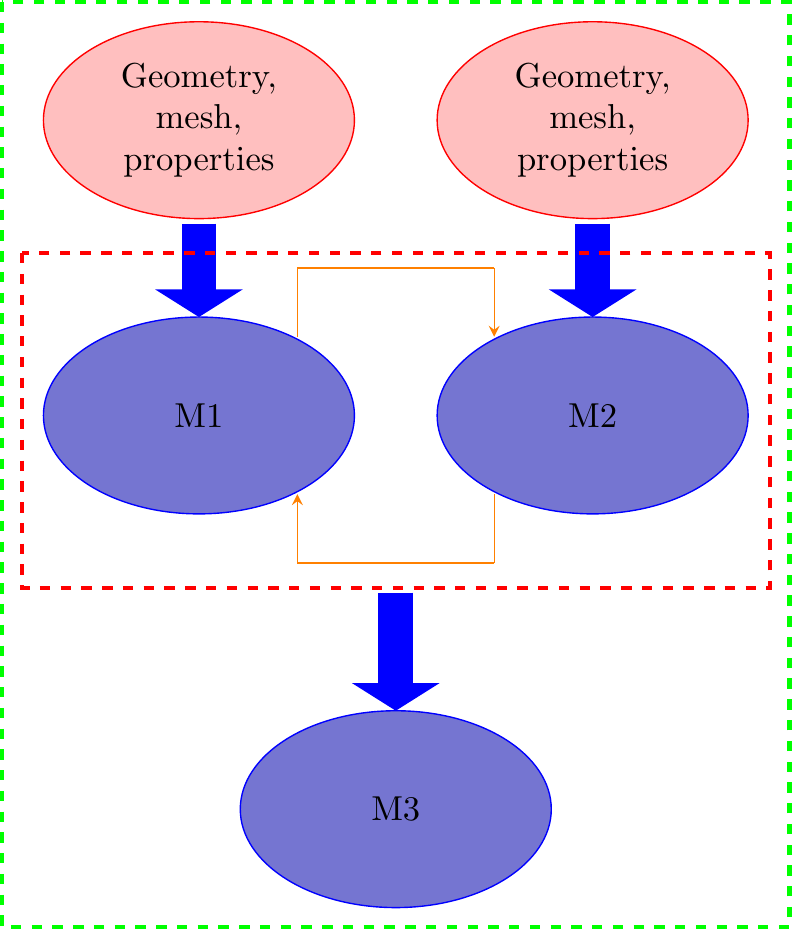}
\\
(a) & (b)
\end{tabular}
\caption{Sequential (a) and Loosely coupled (b) workflow templates using MODA diagrams.} 
\label{Fig2}
\end{figure}

%%%%%%%%%%%%%%%%%%

\section{MuPIF’s data interfaces}
MuPIF abstracts from any particular data representation. This is achieved by representing data entities as objects and defining the standardized interfaces for data types. By analyzing the problem domain, it is possible to identify a set of canonical operations required for each data type. These operations constitute the data type interface. The interface makes it possible to hide the internal representation and data structure for a particular instance. As an example, consider a spatial field data type, representing a variable with spatial distribution. It can be represented in a number of different ways, including as a mathematical formula as a function of spatial coordinates or as an interpolated set of spatially distributed point values. For a spatial field type, the canonical operations include methods for evaluating field value and its derivatives at a given position.
The concept of data type abstraction does not require models to interpret the raw data structure of a data type instance; instead, they can rely on an instance abstract interface to interpret the data. As a result, the concept of abstract data types allows for abstraction from a particular representation of data, enabling the natural support of different data storage formats. One can easily create tailored data representations targeted for a specific purpose and simply plug them into workflows without breaking anything in the software. Typically, a data provider determines the actual representation.
This approach significantly reduces implementation work on the model side, since it is no longer necessary to interpret raw data for data instances—the methods are part of the data instances. There is no need for data transformations, and this can reduce induced numerical errors.
The data interfaces not only make it possible to abstract from an internal data structure, but they also enable abstraction from actual stored data. Data can be located in memory, a file, a database, or even be distributed over the network. To summarize this section, MuPIF’s design is focused on the standardization of data interfaces rather than on the standardization of data structures.

\section{MuPIF’s distributed design}
Complex simulations are generally resource and time-demanding. Distributed and parallel computing aims at alleviating such demands. A common feature of distributed and parallel environments are distributed data structures and concurrent processing on distributed processing nodes. This brings with it an additional level of complexity that must be addressed.
\begin{figure}[h!]
\centerline{\includegraphics[width=0.9\linewidth]{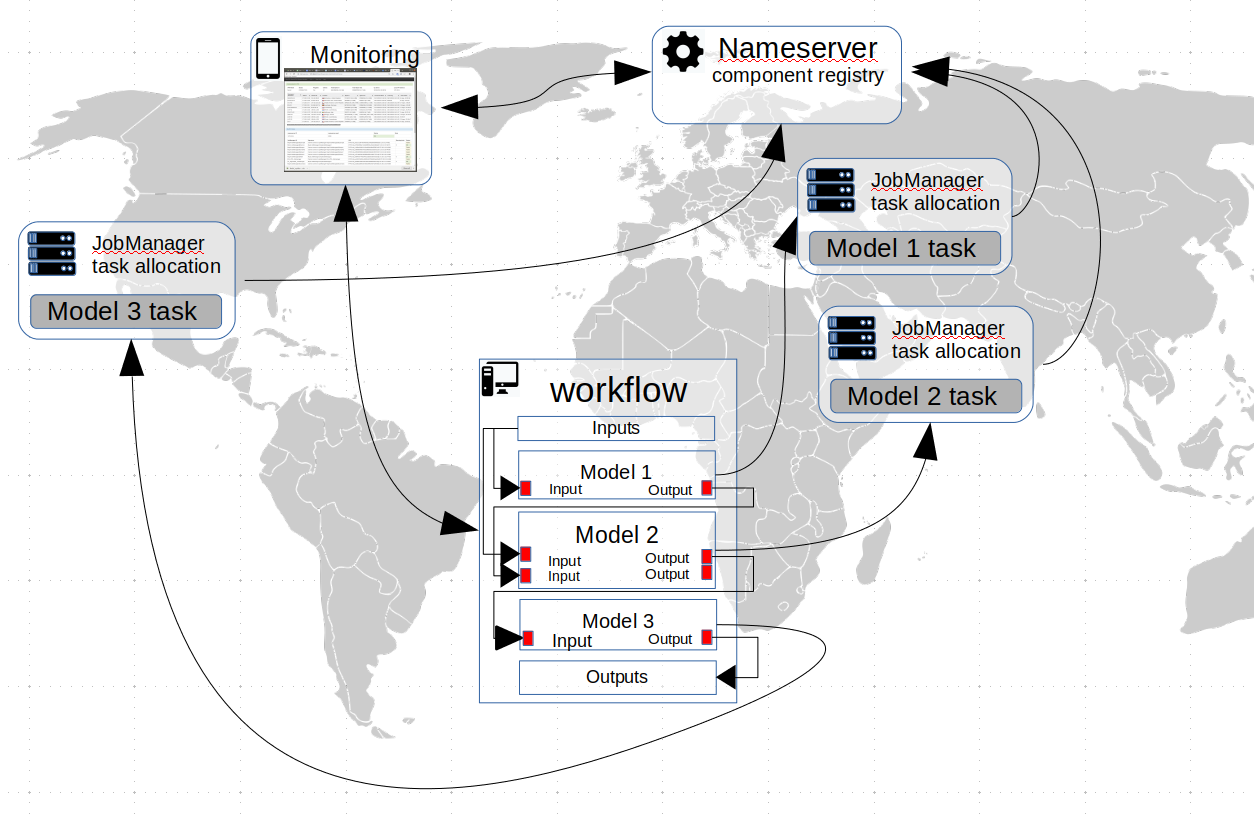}}
\caption{Distributed MuPIF infrastructure.} 
\label{Fig:MuPIF_DistriburedInfrastructure}
\end{figure}
MuPIF provides a communication layer that enables local and network distributed objects to be treated in the same way, hiding the details of distributed network communication from the user. The communication layer is built upon a transparent distributed object system fully integrated into Python 3 using the Pyro4 module \cite{[25]}. The module takes care of network communication between objects when they are distributed over different machines on the network, hiding all socket programming details. One simply calls a method on a remote object as if it were a local object—the use of a remote objects is almost transparent and is achieved by the introduction of so-called proxies. A proxy is a special kind of object that acts as the actual remote object. Proxies forward calls to remote objects and pass results back to the calling code. In this way, there is no difference between simulation workflows for local or remote objects such as models or data entities, except during initialization, when instead of creating local objects one has to connect to their remote counterparts. Such transparent communication makes it possible to decouple development and testing of simulation workflows, which can be performed locally, and it facilitates workflow deployment and execution using distributed resources.
The distributed infrastructure, see Figure 3, also has several utility services. MuPIF provides a remote object registry (\textit{Nameserver}), a sort of “yellow pages” service that enables registration and lookup of objects based on name and object metadata. 
\begin{figure}[h]
\begin{center}
\centerline{\includegraphics[width=1\linewidth]{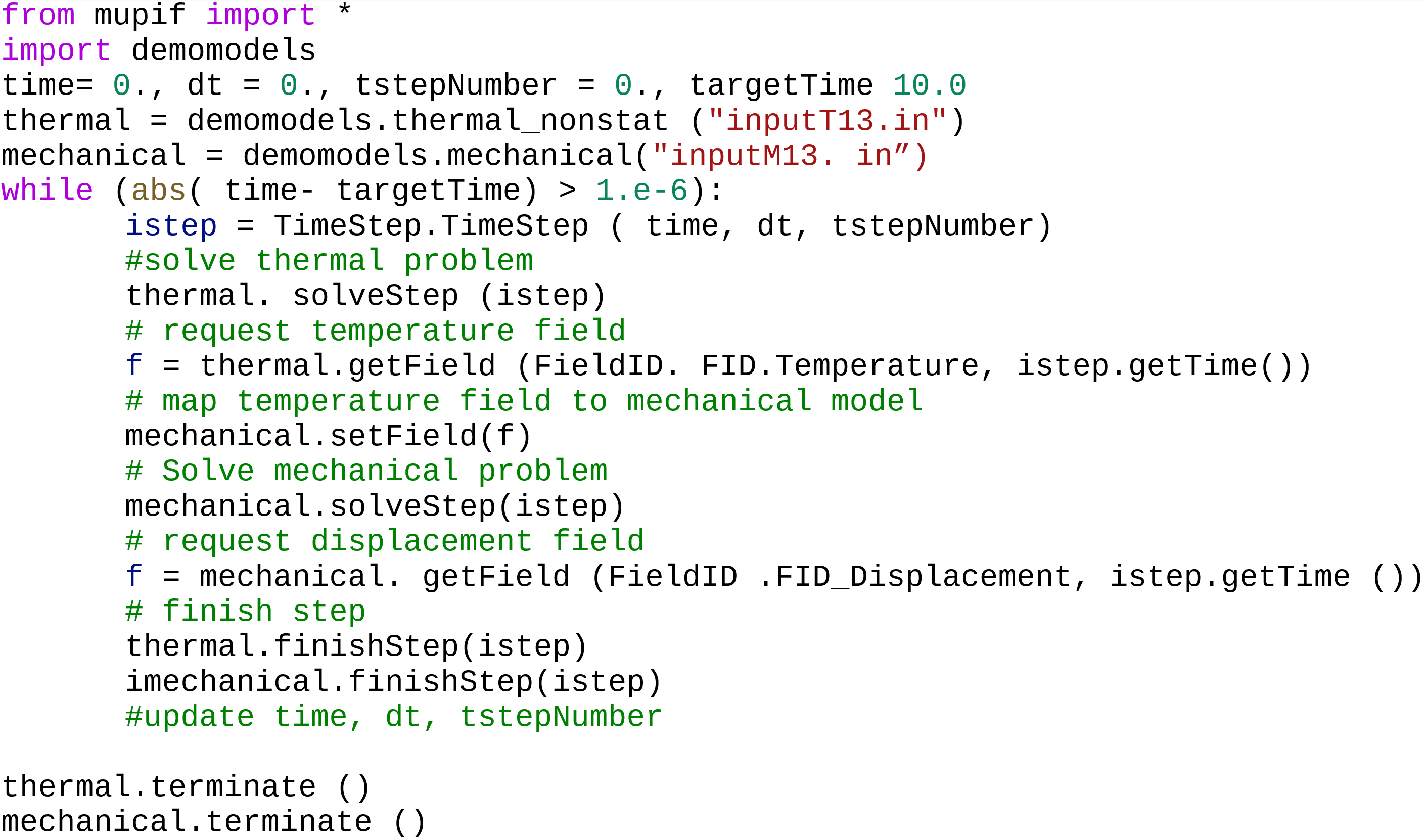}}
\end{center}
\caption{Example of MuPIF simulation workflow representation in Python.} 
\label{Fig:MuPIF_SimulationWorkflow}
\end{figure}
The \textit{JobManager} interface controls resource allocation, defined by a corresponding abstract class. It defines services for allocation and execution of individual model instances and monitoring services. The \textit{JobManager} interface is a Python object like all other platform components. Typically, a unique instance of \textit{JobManager} is created for each application server and registered on the nameserver to make it accessible. This facilitates access to \textit{JobManager}’s services using the same remote object technology as other components and comprises the resource allocation part of a simulation scenario. Typically, a simulation scenario (workflow) first establishes a connection to the platform nameserver, which is used to lookup necessary \textit{JobManagers}. Individual \textit{JobManagers} are subsequently requested to create individual application instances that are remotely accessible by local proxy objects.

\begin{figure}[h]
\centerline{\includegraphics[width=1 \linewidth]{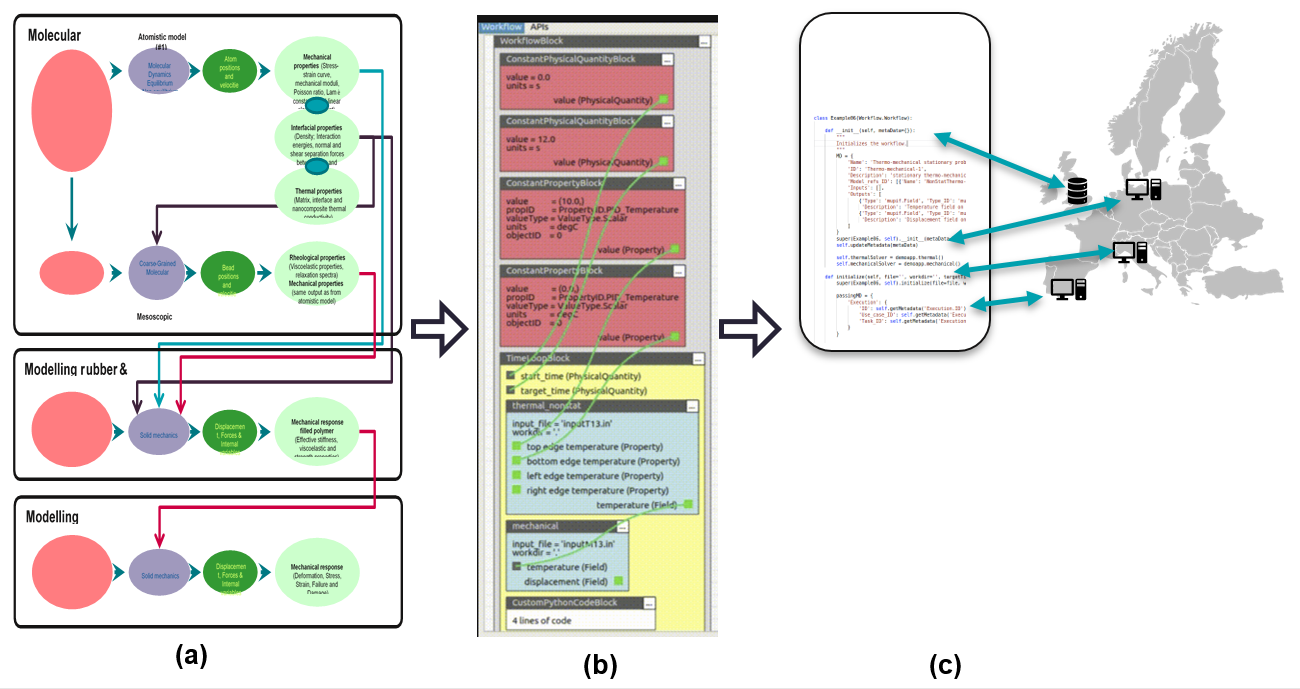}}
\caption{From MODA diagram to executable MuPIF workflows using a graphical workflow editor.} 
\label{Fig:MuPIF_ModaDiagram}
\end{figure}
A virtual private network (VPN) with encryption and authorization services can be used to guarantee communication security.
%%%%%%%%%%%%%%%%%%%%%%%%%%%%%%%%%%%%%%%%%%%%%%%%%%%%%%%%%%%%%%%%%%%%
\section{Simulation workflow and Graphical workflow editor}
The simulation workflow is a computational recipe designed to provide required outputs by combining different models and data sources, see~Figure~\ref{Fig:MuPIF_SimulationWorkflow}. In MuPIF, workflow is represented as a class derived from the \textit{Workflow} class, further derived from the \textit{Model} class. It provides services related to how to get/set data entities and how to execute a workflow. It is a Python script built on top of the MuPIF classes. When developing a workflow, one typically starts from a MODA diagram, defining the coupling/linking of individual models and data into the unique workflow. The translator can either use a graphical workflow editor, allowing workflow to be set up using the graphical editor to generate an executable representation of the workflow, or the executable representation can be written directly on top of the MuPIF platform and its services.

For many users, the need to implement simulation workflows in Python may present an adoption barrier. To alleviate this situation, the MuPIF platform comes with a graphical workflow editor. This editor allows a simulation workflow to be created by combining and linking graphical blocks representing control structures (such as time loops and if/then/else blocks), individual models, and data sources, see Figure \ref{Fig:MuPIF_ModaDiagram}. It is built on the top of abstract data and execution model that is translated into the workflow representation in Python. The data model defines the data dependencies between blocks, and the execution model defines how to generate the code. The design is again object-oriented; all building blocks must implement standardized interfaces, which makes it easy to plug-in new custom building blocks. The workflow editor relies on the metadata provided by the models and data entities to generate input/output data slots automatically and to perform many consistency checks during workflow composition. The workflow editor is written in the JavaScript language and can be run in any web browser from the internet. The editor allows workflow data and execution models to be saved in JSON, makes it possible to generate workflow representation in Python, and facilitates direct execution of a workflow from the editor.
%%%%%%%%%%%%%%%%%%%%%%%%%%%%%%%%%%%%%%%%%%%%%%%%%%%%%%%%%%%%%%%%%%%
\section{EMMO support in MuPIF}
\begin{figure}[h!]
\begin{center}
\centerline{\includegraphics[width=0.95\linewidth]{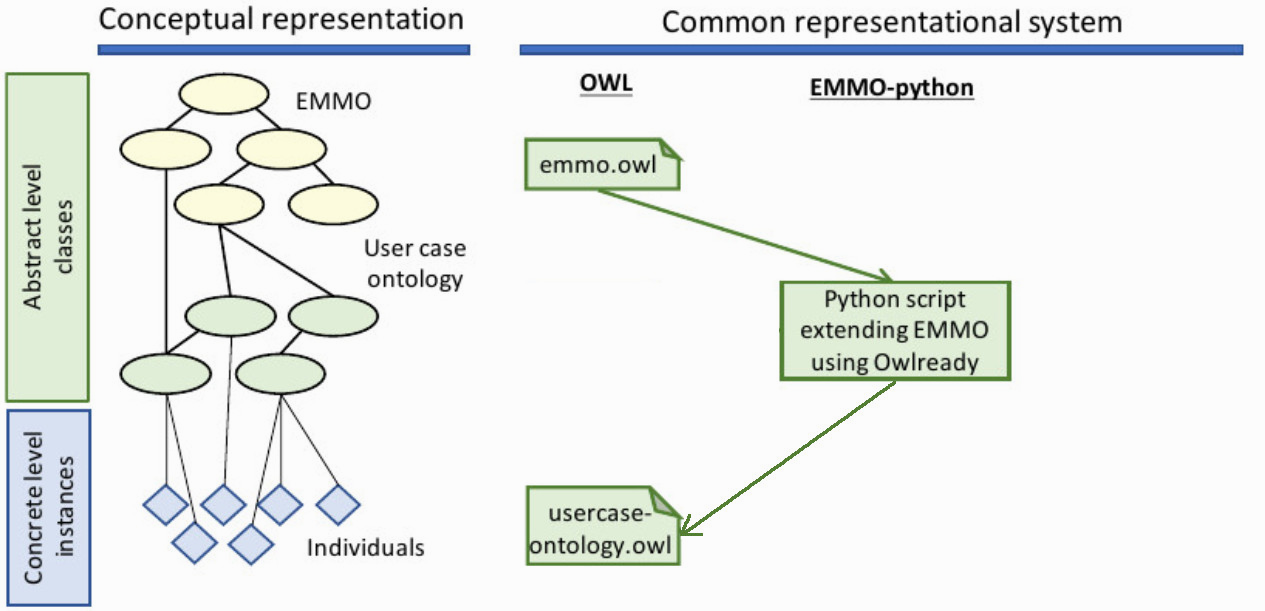}}
\end{center}
\caption{Example of MuPIF integration with EMMO representation ontology.} 
\label{Fig:MuPIF_Ontology}
\end{figure}
The European Materials Modelling Council released recently the European Materials \& Modelling Ontology (EMMO) \cite{[20]}. Its potential is seen as being the semantic connection between models and data, replacing traditional syntactic methods. 

EMMO enables:
\begin{itemize}
    \item Vertical interoperability by extending a model with ontological EMMO classes and relations.
    \item Horizontal interoperability by using an ontological approach to connect different models into a simulation workflow.
\end{itemize}
EMMO defines basic abstract classes and object properties that can later be extended for each particular implementation. MuPIF’s approach for supporting EMMO is to create a user case ontology stemming from the EMMO Owlready2 Python module \cite{[26]}. This provides a natural method for representing the ontology in the form of Ntriples (RDF/XML or OWL/XML format), for creating new classes and individuals in an existing structure, for automatic classification, and for enabling querying tools.

Finally, MuPIF can map its data to ontology instances. This approach is sketched schematically in Figure \ref{Fig:MuPIF_Ontology}, where materials, properties, and input/output data are individuals derived from abstract classes.

In the future this mapping will be natively supported by individual MuPIF classes assisted by the metadata information.

\section{MuPIF database layer and Workflow manager}
Simulations can take advantage of a database system not only for storing simulation data but also for storing a context, making full traceability of simulations possible. The MuPIF platform is designed to be independent of any particular database system. It comes with its own storage solution built on the top of the Mongo Database \cite{[27],[28]}. Data about use cases, workflows, workflow executions, and input and output cards for workflow executions can be stored. The dynamic character of the Mongo database, which is a non-SQL system, makes it possible to design a fully dynamic database solution, allowing complex data structures and their metadata for different simulation workflows to be stored. The database layer comes with:
\begin{itemize}
    \item A workflow scheduler: A module for automatically performing scheduled workflow executions.
    \item A representational state transfer \cite{[29]} (REST) interface: Data from the database can be queried from this interface, new executions can be created, their input parameters can be defined, workflow executions can be scheduled, and results can be inspected. The REST API, see Table \ref{Tab:DB}, facilitates interaction with the database and MuPIF on an abstract level. This was used to serve the Business Decision Support System during the Composelector project (see the “Examples” section for more details) and can also serve various user applications built on top of the database layer.
\end{itemize}

\begin{table}[]
\footnotesize
\begin{tabular}{|p{5.cm}|p{7.8cm}|}
\hline
{\textbf{Service}}  & {\textbf{Description}}                    \\ \hline
/status                                 & Returns the status of MuPIF DB                                                                                                                \\ \hline
/usecases                               & Returns list of defined use ceases                                                                                                            \\ \hline
/usecases/ID                            & Returns details of use case with 			given ID                                                                                                  \\ \hline
/usecases/ID/workflows                  & Returns list of workflows 			available for use case ID                                                                                        \\ \hline
/workflows/ID                           & Returns details of workflow 			identified by ID                                                                                               \\ \hline
/workflowexecutions/init/ID             & Initialize (schedules) execution 			of workflow with ID, returns new workflow execution ID (WEID)                                             \\ \hline
/workflowexecutions/WEID                & Show status of workflow execution 			with WEID                                                                                                \\ \hline
\makecell[l]{/workflowexecutions/WEID/\\inputs}         & Returns inputs for workflow 			execution WEID                                                                                                 \\ \hline
\makecell[l]{/workflowexecutions/WEID/\\outputs}        & Returns outputs for workflow 			execution WEID                                                                                                \\ \hline
\makecell[l]{/workflowexecutions/WEID/set?\\NAME=value} & \makecell[l]{ Sets input parameter for workflow 			execution WEID \\identified by NAME. The format of value is string \\with format depending on input type.} \\ \hline
\makecell[l]{/workflowexecutions/WEID/get?\\NAME}       & \makecell[l]{Returns value of output parameter 			identified by \\ NAME for workflow execution WEID.}         
\\ \hline
/executeworkflow/WEID                   & Schedules execution  of workflow 			execution WEID.                                                                                           \\ \hline
\end{tabular}
\label{Tab:DB}
\caption{Core MuPIF DB REST Services.}
\end{table}

\section{MuPIF’s monitor
}
\begin{figure}[h!]
\begin{center}
\centerline{\includegraphics[width=0.99\linewidth]{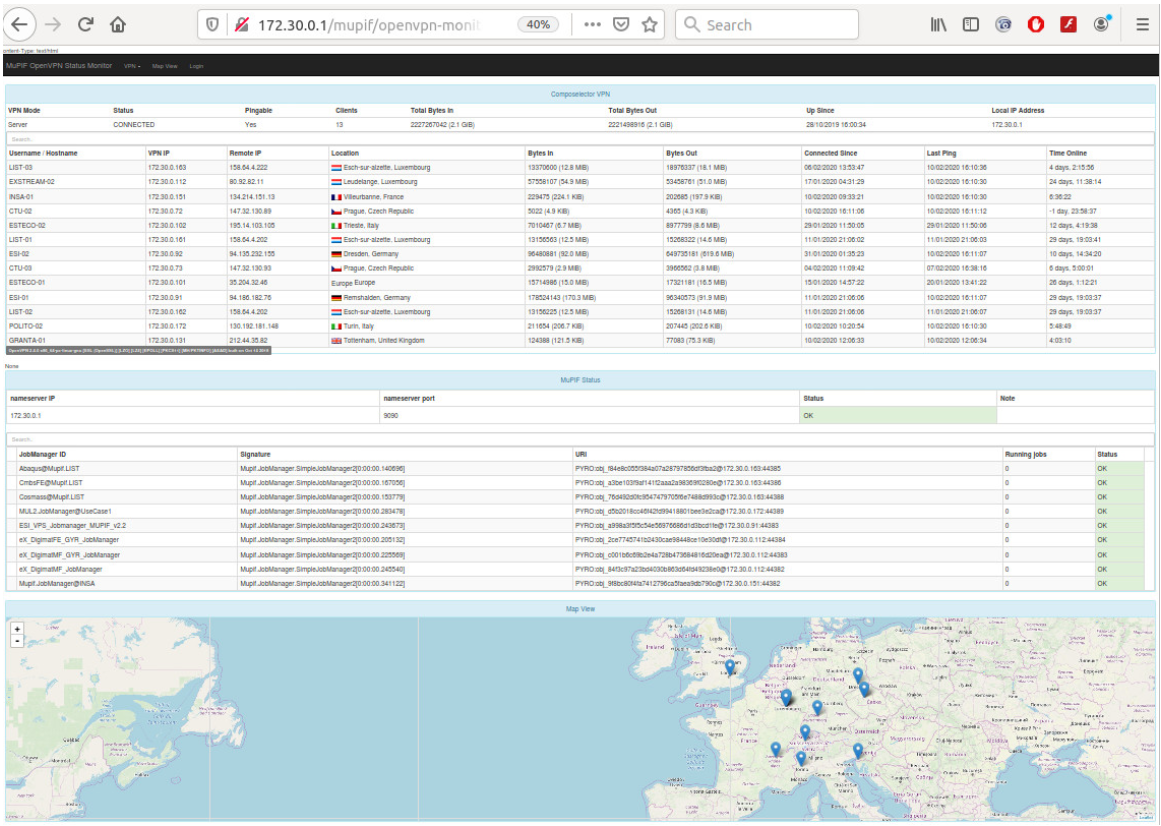}}
\end{center}
\caption{Screenshot of MuPIF monitoring tool.} \label{Fig:MuPIF_Monitor}
\end{figure}
Platform status and operation can be monitored online using a web-based monitoring tool based on the OpenVPN monitoring tool, see Figure \ref{Fig:MuPIF_Monitor}. The tool makes it possible to monitor the status of a VPN network, connected clients, and their statistics (locations, IP address, transferred data volume, and so on). With it, the state of a name server can be visualized and  a directory of registered objects (structured by the object type) can be displayed together with status. For remote resources, the number of running jobs is shown, among other things. The monitoring tool also comes with an administrative interface that allows specific administrative tasks to be performed on platform components using the web interface.

%%%%%%%%%%%%%%%%%%%%%%%%%%%%%%%%%%%%%%%%%%%%%%%
\section{Composelector project}
\begin{figure}[h!]
\begin{center}
\centerline{\includegraphics[width=0.82\linewidth]{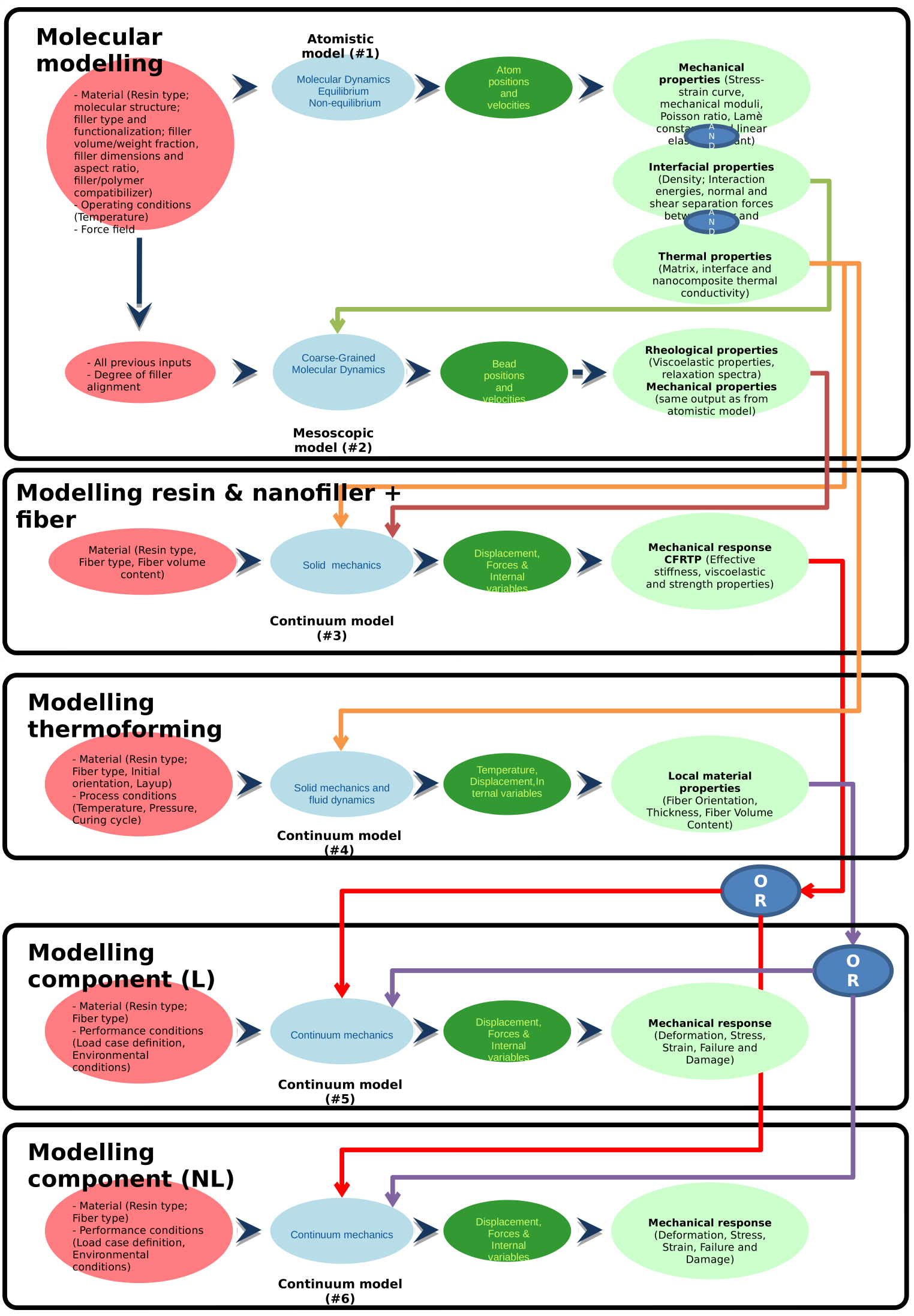}}
\end{center}
\caption{From MODA diagram to executable MuPIF workflows using a graphical workflow editor.} 
\label{Fig:MuPIF_Moda}
\end{figure}
\begin{figure}[h!]
\begin{center}
\centerline{\includegraphics[width=0.99\linewidth]{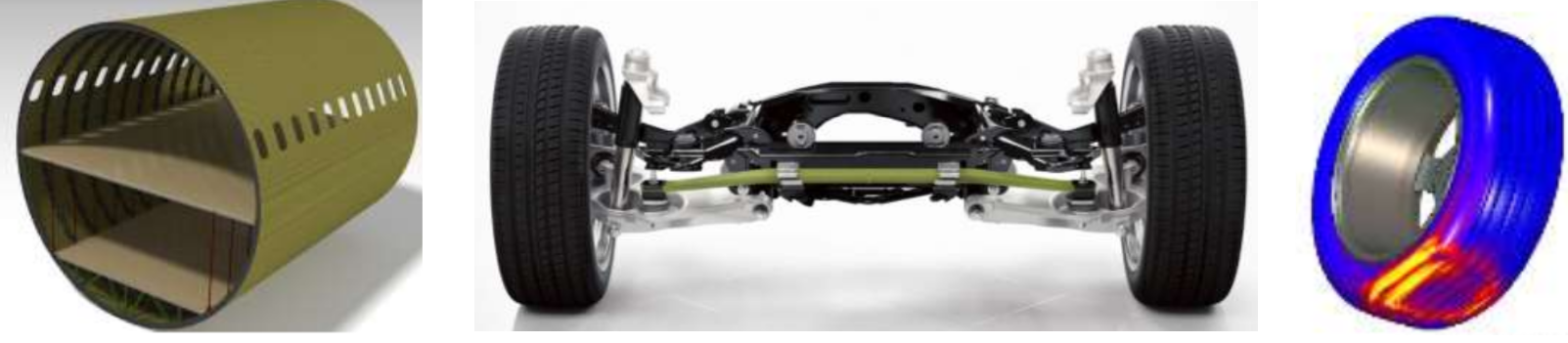}}
\end{center}
\caption{Composelector use cases: Composite airplane frame, composite leaf-spring and innovative tire design.} 
\label{Fig:MuPIF_UseCase}
\end{figure}
\begin{figure}[h!]
\begin{center}
\centerline{\includegraphics[width=1\linewidth]{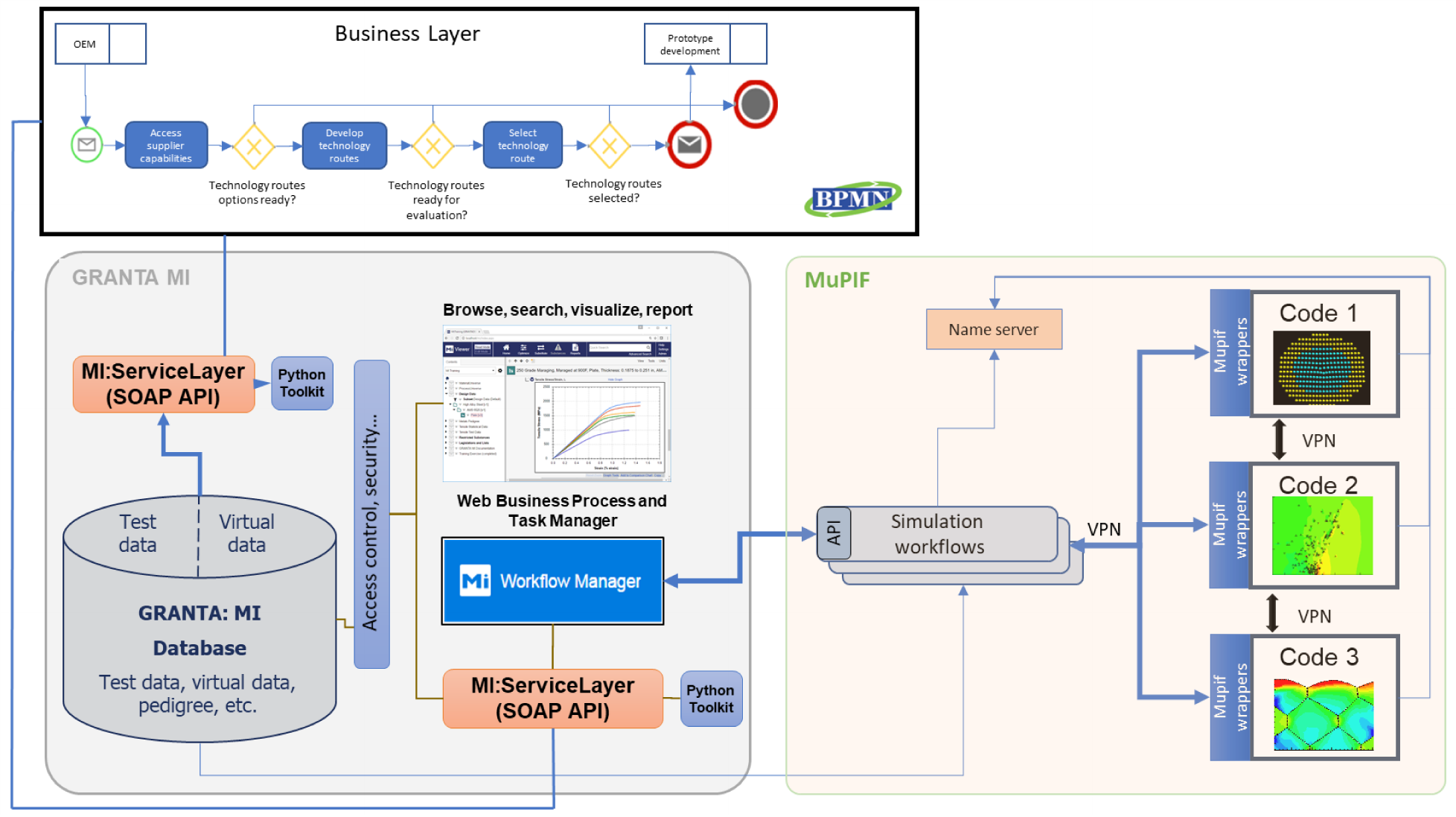}}
\end{center}
\caption{Composelector BDSS schema.} 
\label{Fig:MuPIF_BDSS}
\end{figure}

At present, the platform is being integrated into a Business Decision Support System (BDSS) in the frame of the EU H2020 Composelector project~\cite{[30]}. The project aims to develop a methodology for designing innovative composite materials with applications in the aerospace industry (design of composite airplane frame, see Figure \ref{Fig:MuPIF_Moda} for modeling workflow) and in the automotive industry (design of composite leaf springs and tires), see Figure \ref{Fig:MuPIF_UseCase}. The complex simulation workflows involve a combination of several simulation tools starting at the atomistic/molecular scale to determine basic material properties, bridging several resolution scales and ending up at the component/structural scale, including commercial and open-source modeling tools distributed and executed across different locations and computing resources around Europe.
The BDSS makes it possible to model business decision process during new, user-driven product development, including different actors (at the business and/or technical level) and to automate decisions based on modeling. The Business Process Model and Notation (BPMN) industrial standard  provides a graphical web based editor based on ESTECO technology \cite{[31]}. It integrates the Granta \cite{[32]} material database to access and store data about materials extended to manage simulation data. MuPIF serves as a simulation platform, where simulation workflows are defined and executed. Figure \ref{Fig:MuPIF_BDSS} illustrates the concept for the Composelector platform.

\section{Application example: Design of composite airplane frame}
\begin{figure}[h!]
\begin{center}
\centerline{\includegraphics[width=0.9\linewidth]{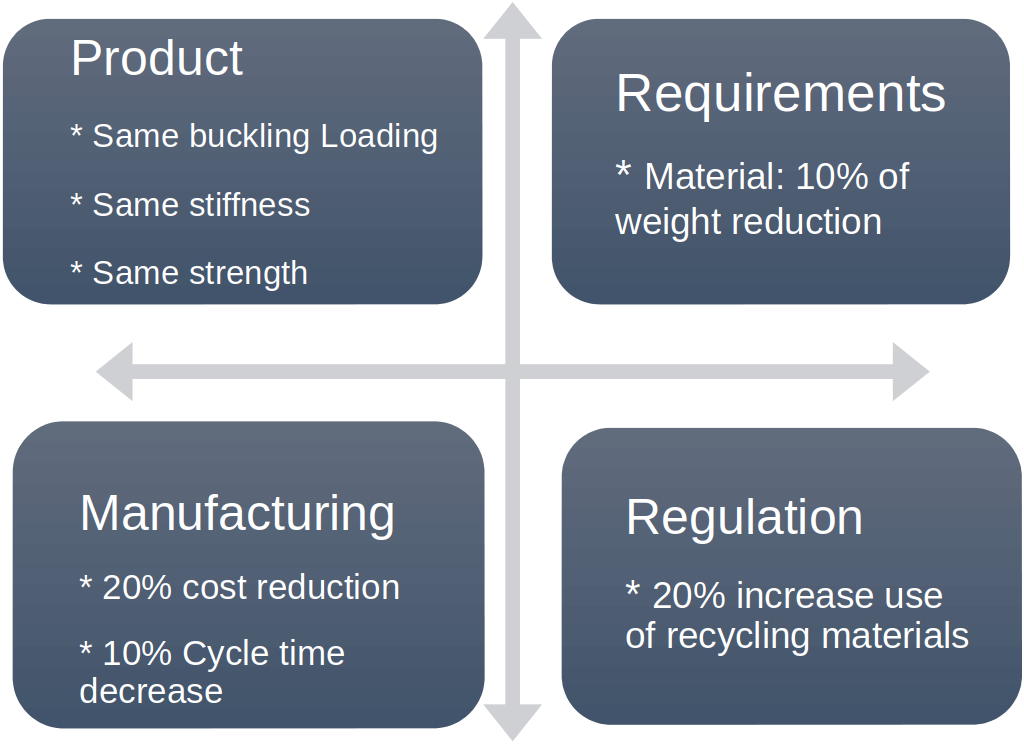}}
\end{center}
\caption{Requirements on user case of design of composite airplane.} 
\label{Fig:Requirements}
\end{figure}

\begin{figure}[h!]
\begin{center}
\centerline{\includegraphics[width=0.95\linewidth]{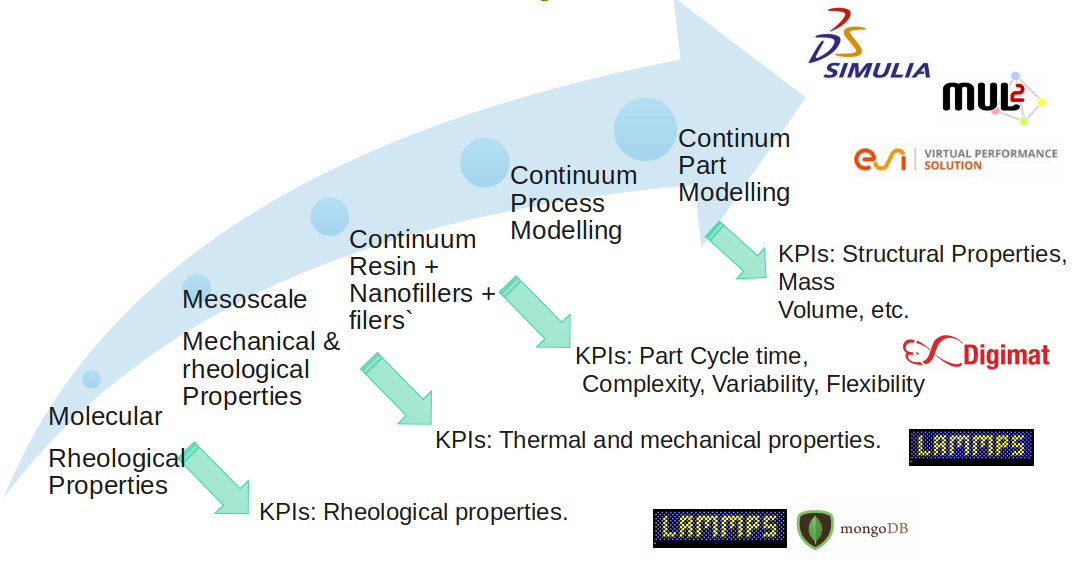}}
\end{center}
\caption{Material KPI extraction.} \label{Fig:KPI}
\end{figure}

In the following section, a use case for a composite airplane frame will be described in more detail to illustrate MuPIF’s role. The strategic objective in this use case is the exploration of options for moving from a thermoset to a thermoplastic material. At the same time, there are several requirements from the product, manufacturing, and regulation points of view,~see~Figure~\ref{Fig:Requirements}. The performance of a design is measured using Key Performance Indicators (KPIs) that follow from the initial requirements. Many parameters, including geometry, have been fixed; only several parameters controlling material composition and processing have not been defined. From the MODA description, 22 different modeling workflows have been identified and implemented in total, starting at different resolution levels (from molecular modeling of the resin to continuum analysis at the component scale using different simulation tools to preform individual modeling tasks, see~Figure~\ref{Fig:KPI}). The individual workflows are summarized in Table~\ref{Tab:SimulationWorkflows} and have been integrated in the database layer and are accessible for use either from the Business Layer interface or the MuPIF web API.

\begin{table}[]
\footnotesize
\begin{tabular}{|l|l|}
\hline
\textbf{ID} & \textbf{Description}                                                                                                                                \\ \hline
A\_1                              & DB-\textgreater MUL2, KPI 1 - 2 (buckling load)                                                                                                                                           \\ \hline
A\_2                              & Digimat-\textgreater MUL2, KPI 1 - 2 (buckling load)                                                        \\ \hline
A\_3                              & Lammps -\textgreater Digimat-\textgreater MUL2, KPI 1 - 2 (buckling load)                                                                                                        \\ \hline
A\_4                              &DB-\textgreater Abaqus, KPI 1 - 2 (buckling load)                                                                                                                            \\ \hline
A\_5                              & Digimat -\textgreater Abaqus, KPI 1 - 2 (buckling load)                                                                                                                                 \\ \hline
A\_6                              & Lammps -\textgreater Digimat -\textgreater Abaqus, KPI 1 - 2 (buckling load)                                                                                                        \\ \hline
A\_10                             &Lammps -\textgreater Digimat, KPI 1 - x (elasticity)                                                                                                                                  \\ \hline
A\_11                             & Digimat, KPI 1 - x (elasticity)                                                                                                                 \\ \hline
A\_12                             & DB-\textgreater Abaqus, KPI 1 - 1 (weight) and KPI 1 - 3 (stiffness)                                                                                                \\ \hline
A\_13                             & Digimat -\textgreater Abaqus, KPI 1 - 1 (weight) and KPI 1 - 3 			(stiffness)                                                                                                            \\ \hline
A\_14                             & Lammps -\textgreater Digimat -\textgreater Abaqus, KPI 1 - 1 (weight) and KPI 1 - 			3 (stiffness)                                                                         \\ \hline
A\_17                             & \makecell[l]{Lammps -\textgreater Digimat -\textgreater Comsol -\textgreater Abaqus, KPI 1 - 1 			(weight) and \\ KPI 1 - 3 (stiffness)}                                               \\ \hline
A\_18                             & \makecell[l]{Lammps -\textgreater Digimat -\textgreater Comsol -\textgreater Abaqus (2x), KPI 1 - 1 			(weight) and \\ KPI 1 - 3 (stiffness)}                                             \\ \hline
A\_19                             & \makecell[l]{DB 			-\textgreater Abaqus (2x), KPI 1 - 1 (weight), KPI 1 - 2 (buckling load),\\ 			KPI 1 - 3 (stiffness) and KPI 1 - 4 (maximum Mises Stress)}                   \\ \hline
A\_20                             & \makecell[l]{DB 			-\textgreater Digimat -\textgreater Abaqus (2x), KPI 1 - 1 (weight), KPI 1 - 2 			(buckling load),\\ KPI 1 - 3 (stiffness) and KPI 1 - 4 (maximum 			Mises Stress) }                    \\ \hline
A\_21                             & \makecell[l]{DB -\textgreater Comsol -\textgreater Abaqus (2x), KPI 1 - 1 (weight), KPI 1 - 2 			(buckling load),\\ KPI 1 - 3 (stiffness) and KPI 1 - 4 (maximum 			Mises Stress)}                       \\ \hline
A\_22                             & \makecell[l]{DB 			-\textgreater Digimat -\textgreater Comsol -\textgreater Abaqus (2x), KPI 1 - 1 (weight),\\
KPI 1 - 2 (buckling load), KPI 1 - 3 (stiffness) and KPI 1 - 4 			(maximum Mises Stress)} \\ \hline
\end{tabular}
\label{Tab:SimulationWorkflows}
\caption{Example of simulation workflows for the design of composite airplane frame.}
\end{table}
%%%%%%%%%%%%%%%%%%%%%%%%%%%%%%%%%%%%%%%%%%%
\section{Preliminary results}
\begin{figure}[h!]
\begin{center}
\centerline{\includegraphics[width=0.6\linewidth]{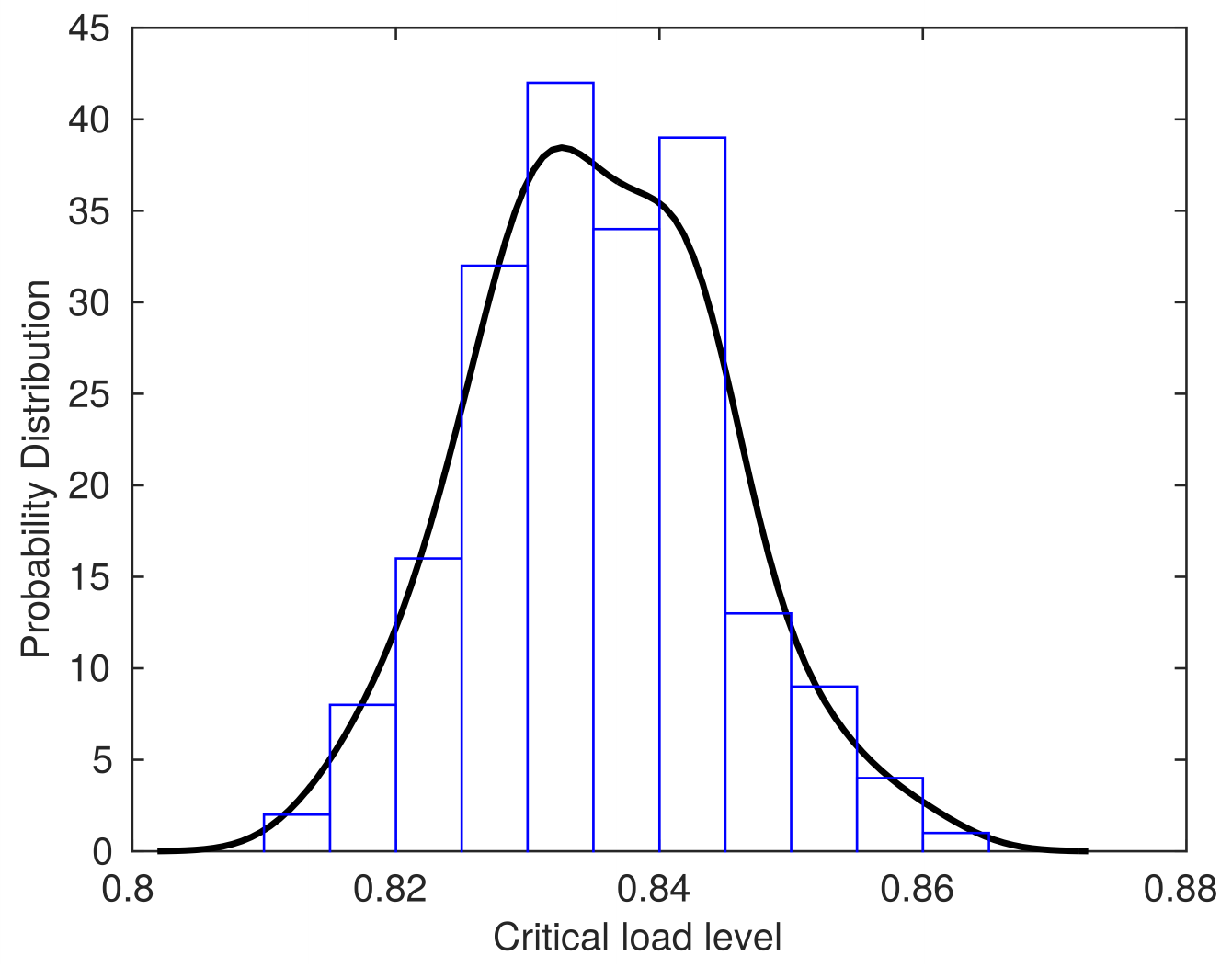}}
\end{center}
\caption{Probability density function of the buckling mode (histogram and fitted density).} 
\label{Fig:MuPIF_BuclingProbability}
\end{figure}
In this section, we present preliminary results for buckling analysis with the uncertainties for a composite air frame as described above. Buckling is a form of instability that leads to the failure of a structure. It occurs when a structure or part of a structure is subject to compression. 
When applied load increases, a structure becomes unstable at some point called the “critical load level.” Uncertainties make it challenging to predict buckling accurately.
In this use case, we consider input uncertainties at the micro-scale level in the material properties of epoxy resin. We notice that any change in material density do not make any difference in the buckling mode. Thus, for this analysis, uncertainties are introduced in the volume fraction, Young’s modulus, and Poisson ratio for the epoxy resin. At the mesoscale, uncertainties are introduced in the stacking sequence of layers (i.e., in layer thickness and the orientation angles). A total of 32 uncertain parameters~(16~orientation angles and 16 layer thicknesses) are considered for the frame. For the~12~layers in the stringer, a total of 24 uncertain parameters (12 thicknesses and 12 angles) are also included. A third order polynomial approximation has been considered to obtain accurate stochastic results. For the sparse polynomial based surrogate, a total of 200 samples using the LHS scheme was considered, see \cite{[33]} for more detail.
Figure \ref{Fig:MuPIF_BuclingProbability} shows the probability distribution of the first buckling mode. In it, the range of the buckling mode due to the input uncertainties can be observed. The probability distribution of the buckling mode is similar to a Gaussian distribution. As for structural stability, the buckling mode provides essential information related to failure, and it is worth knowing its sensitivity with respect to input parameters in order for the resultant engineering design to be safe and reliable.

\begin{figure}[h!]
\begin{center}
\centerline{\includegraphics[width=0.9\linewidth]{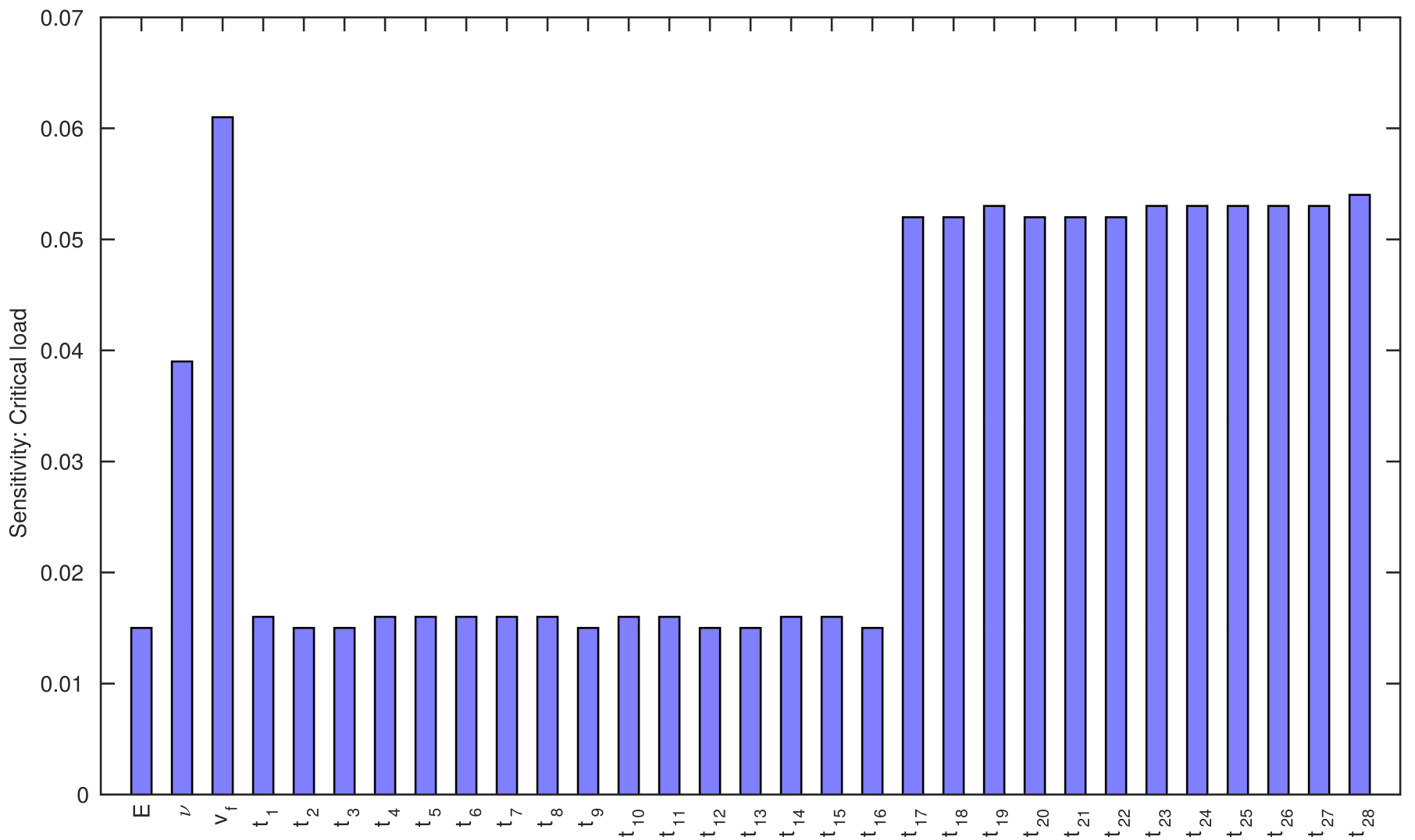}}
\end{center}
\caption{Sensitivity of input parameters on the buckling mode.} 
\label{Fig:MuPIF_BuclingSensitivity}
\end{figure}

In Figure \ref{Fig:MuPIF_BuclingSensitivity}, the sensitivity information is also provided. Sensitivity information on the buckling mode is shown for the input parameters. It can be noticed that from the micro-scale, the volume fraction of the materials used in the composite is the most sensitive parameter for the buckling analysis. A small change in the volume fraction can make a huge impact on the failure of the structure. Young's modulus of the epoxy resin exhibits minor effects on the buckling mode. The sensitivity in the Poisson ratio is also a very dominating parameter on the buckling mode. The thicknesses of the composite layers in the laminates exhibit a good impact on the buckling mode. The buckling mode is not sensitive to changes in the orientation angles of the composite layers. Figure \ref{Fig:MuPIF_BuclingSensitivity} also provides sensitivity information regarding the buckling mode for certain input parameters. Starting at the micro-scale, the volume fraction for the materials used in the composite is the most sensitive parameter for buckling analysis. A small change in the volume fraction can make a tremendous impact on the failure of the structure. The Young’s~modulus~of the epoxy resin exhibits minor effects on the buckling mode. The sensitivity in the Poisson ratio is also a very dominating parameter for the buckling mode. The thicknesses of composite layers in the laminates exhibit a substantial impact on the buckling mode. The buckling mode is not sensitive to changes in the orientation angles of the composite layers.
\section{Conclusions}
This article presents the design of MuPIF, a distributed, open-source simulation platform and highlights key design features including the object-oriented concept of standardized interfaces for models and data types, the concept of hierarchical workflows, the distributed object system, and the integrated database solution. These features make the platform unique. MuPIF can be incorporated into complex scenarios and distributed environments and can handle multiple data formats. A use case for the design of a composite airplane frame illustrates use and operation of the platform, which has matured into a fully operational system able to be used in different settings. MuPIF’s standardized REST interface makes it possible to integrate it with various material design platforms and workflows.

\section*{Acknowledgement}
The authors would like to acknowledge the support of EU H2020 COMPOSELECTOR project (GA no: 721105). 

%% The Appendices part is started with the command \appendix;
%% appendix sections are then done as normal sections
%% \appendix

%% \section{}
%% \label{}

%% If you have bibdatabase file and want bibtex to generate the
%% bibitems, please use
%%
%%  \bibliographystyle{elsarticle-harv} 
%%  \bibliography{<your bibdatabase>}

%% else use the following coding to input the bibitems directly in the
%% TeX file.

\end{document}